\documentclass[sigconf,review, 10pt]{acmart}
 \renewcommand\footnotetextcopyrightpermission[1]{}

\newcommand{\oursys}{WixUp}

\usepackage{algorithm}
\usepackage{algpseudocode}

\usepackage{subcaption}

\begin{document}

\title{\oursys: A General Data Augmentation Framework for Wireless Perception in Tracking of Humans}

\begin{abstract}
Wireless sensing technologies, leveraging ubiquitous sensors such as acoustics or mmWave, can enable various applications such as human motion and health tracking. However, the recent trend of incorporating deep learning into wireless sensing introduces new challenges, such as the need for extensive training data and poor model generalization. As a remedy, data augmentation is one solution well-explored in other fields such as computer vision; yet they are not directly applicable due to the unique characteristics of wireless signals. Hence, we propose a custom data augmentation framework, \oursys, tailored for wireless human sensing. Our goal is to build a generic data augmentation framework applicable to various tasks, models, data formats, or wireless modalities.
Specifically, \oursys{} achieves this by a custom Gaussian mixture and probability-based transformation, making any data formats capable of an in-depth augmentation at the dense range profile level.
Additionally, our mixing-based augmentation enables unsupervised domain adaptation via self-training, allowing model training with no ground truth labels from new users or environments in practice.
We extensively evaluated \oursys{} across four datasets of two sensing modalities (mmWave, acoustics), two model architectures, and three tasks (pose estimation, identification, action recognition). \oursys{} provides consistent performance improvement (2.79\%-84.25\%) across these various scenarios and outperforms other data augmentation baselines.

\end{abstract}


%
%





\maketitle

\section{Introduction}

Recent advancements in wireless sensing have enabled a broad spectrum of applications, ranging from human tracking and health monitoring to large radar systems for self-driving and robotics. It includes various wireless sensing modalities using ubiquitous sensors, such as mmWave, acoustics, WiFi, etc.
These technologies offer several advantages over traditional camera-based perception systems. Wireless signals are not restricted by light conditions or line-of-sight occlusions, and they provide better privacy protection compared to cameras. 
Typically, a wireless sensing system emits custom signals, such as frequency-modulated continuous wave (FMCW) from the transmitter~\cite{lovescu2020fundamentals}; these signals are reflected by various objects and people in the environment. Then, by analyzing the reflections recorded by the receiver, it can extract information for localization or motion tracking.

The recent emergence of utilizing deep learning in wireless sensing greatly improves their tracking capability~\cite{zhao2018rf, zhao2018through}, but also brings forth new challenges. Firstly, there is a significant demand for collecting extensive training data that requires considerable resources and effort.
Secondly, the trained model often exhibits poor generalization when deployed in new scenarios, such as new users or environments.
Moreover, when the raw wireless signal is processed into point clouds, it exhibits severe sparsity issues that affect the training data quality. Consequently, downstream tasks face limitations in accuracy and struggle to address complex tasks, such as flow estimation. For instance, Fig.~\ref{fig:skeleton} demonstrates two samples of the processed mmWave as point clouds, which appear sparse and barely resemble a coherent human skeletal structure; while the depth camera provides dense point clouds. The disparity is owing to the inherent limitations of wireless technology, where its low granularity is insufficient for multi-part tracking.

To address these challenges, data augmentation is an effective solution, proven in many fields of machine learning~\cite{morningstar2024augmentations, zhang2017mixup}. There are well-explored methods for augmenting 2D or 3D data like images~\cite{zhang2017mixup} and Lidar point clouds~\cite{xiao2022polarmix}, broadly categorized into global augmentation and local augmentation~\cite{xiao2022polarmix}. 
1) Global augmentation performs random scaling, flipping, and rotation on 2D images or 3D Lidar point clouds to transform individual data. However, they fail to augment local structures and ignore the relevance between elements within individuals.
2) Instead, local augmentation techniques tend to mix two data to generate new samples. Mixup~\cite{zhang2017mixup} uses convex combination to mix data, \colorblock{theoretically proved that mixing-based data augmentation is effective for generalization}; ~\cite{yun2019cutmix, xiao2022polarmix} crop\&paste local structure of patch or an object from one data to another.
Additional benefits of mixing-based data augmentation are that, beyond improving model performance in supervised learning, it helps perform unsupervised domain adaptation via self-training\cite{xiao2022polarmix}, which reduces labeling efforts.
Nevertheless, mixing-based augmentation requires designing a domain-specific patch/object selection algorithm.

Some current wireless research has adopted straightforward global augmentation methods, such as rotation or Gaussian noise to classify RF signal types~\cite{huang2019radiomodule, chen2023radiowavelet, clark2021training, blindsignal2019}, and shifting the range or initial phase for perception~\cite{acousticdataset2023}. 
However, mixing-based local augmentation remains unexplored, due to its complexity in incorporating domain-specific representations.
Also, the above global methods are confined to specific tasks or models and often lack a comprehensive evaluation.

Besides, a recent trend involves leveraging generative models to simulate and synthesize wireless data, sharing similar motivation with data augmentation; but they are also limited to certain tasks or data. In brief, some cross-modality synthesis work use videos\cite{ahuja2021vid2doppler, 2023synthesizemili} or text prompts\cite{chen2023rfgenesis} to generate RF signals; but they are constrained by limited source video or inadequate details conveyed by text description. Besides, \cite{zhao2023nerf2, chi2024rfdiff} use NeRF or Diffusion to simulate RF signals, only aiding static tasks like localization and channel estimation or simple gesture classification; intensive labeled data are also required to train the model first.
Overall, generative methods demand non-trivial effort to train an extra model and can hardly generate out-of-distribution data.

In this work, we aim to propose a mixing-based data augmentation framework, that is training-free and particularly good at increasing data diversity or even closing domain gaps via self-training~\cite{xiao2022polarmix}. 
It could support multiple human tracking tasks, model architectures, and wireless sensing modalities (like RF or acoustics). Furthermore, beyond improving model performance, it could also reduce labeling efforts by unsupervised domain adaptation via self-training. In practice, this enables training the model with new user/environment data without any labels. These benefits make our framework significantly more practical for wireless sensing than either conventional global augmentation or generative simulation. \colorblock{Ablation studies also justify the effectiveness and rationality of the approach.}

However, this goal poses two primary challenges:
1) We need a mixing algorithm tailored to the features of wireless signals. Unlike 2D images or Lidar point clouds, wireless raw data is less interpretable, rendering many conventional augmentation strategies not applicable, like flipping. Even after processing the raw data into point clouds, the noise and the sparsity issue we discussed in Fig.~\ref{fig:skeleton} make existing 3D data augmentations insufficient.
2) To perform a comprehensive evaluation of our method on diverse real-world data and tasks, we need access to multiple high-quality datasets. Excitingly, in the realm of mmWave sensing, we have started seeing trends of open-sourcing data recently~\cite{cui2024milipoint, yang2024mm, marsdataset2021}, but they do not provide the unprocessed raw data. The processed point clouds are lossy information, because of the constant false alarm rate (CFAR) filtering as part of the standard FMCW processing pipeline. 

Thereby, we propose \oursys, a range-profile-level data augmentation tailored for the unique characteristics in wireless perception, incorporating the Gaussian mixture to solve the lossy point cloud issue.
1) To elaborate, we employ a custom data processing pipeline to transform any format of raw data into the lossless space of range profile, involving simulating irreversible transformation of CFAR. This is based on representing range profile as a Gaussian mixture of various ranges. Therefore, mixing two data points for data augmentation is equivalent to the intersection of two Gaussian mixtures. Additionally, probability-based methods can further bootstrap the augmented data; embodying spherical angles over ranges can yield 3D outputs. 
2) Finally, by framing a unified code base, we meticulously investigate the generalizability of our data augmentation framework across four datasets involving two sensing modalities (mmWave and acoustics), two model architectures, and three tasks, with a focus on human perception because this topic has most open research resource available to make this study ready. Furthermore, we demonstrate its efficacy in unsupervised domain adaptation via self-training, across unseen environments and users.
Experiment results show that our data augmentation framework exhibits a consistent improvement across different evaluation experiments over baselines.

In summary, our main contributions are:
\begin{itemize}
    \item We propose a custom mixing-based (local) data augmentation technique for wireless sensing that preserves characteristic representations for in-depth augmentation and handles imperfect data types.
    \item Our framework could incorporate self-training to reduce labeling efforts by unsupervised domain adaptation via self-training. This enables training the model with no labels from the new environments or users.
    \item We conduct comprehensive experiments in a unified setup to demonstrate \oursys 's generalizability across tasks, model architectures, datasets, and sensing modalities. 
\end{itemize}

\section{Related work}
\textbf{Data augmentation techniques} (DA) is well-explored for 2D images~\cite{he2016deep, tan2019efficientnet, ren2015faster} and 3D Lidar point clouds~\cite{xiao2022polarmix}, since collecting and annotating training data is labor-intensive and time-consuming.
DA of 2D images typically refers to global augmentation techniques such as random cropping~\cite{krizhevsky2012imagenet, szegedy2015going}, scaling~\cite{simonyan2014very, szegedy2015going}, erasing~\cite{zhong2020random}, and color jittering~\cite{szegedy2015going}, which aim to learn transformation invariance for image recognition tasks. In contrast, local augmentation techniques, such as mixup~\cite{zhang2017mixup} and CutMix~\cite{yun2019cutmix}, generate new training data through various mixing operations.
3D DA has garnered attention recently, which also adopts similar methods such as scaling, rotation, and translation; Recent studies attempt to augment local structures of point clouds~\cite{li2020pointaugment, sheshappanavar2021patchaugment, kim2021point} and introduce concepts for object-level augmentation~\cite{chen2020pointmixup, lee2021regularization} and scene-level~\cite{xiao2022polarmix} in self-driving.
However, conventional DAs are unsuitable for addressing the unique characteristics of wireless sensing data, as we discussed above, necessitating the development of customized approaches.

\textbf{Data augmentation for labeling efficiency.}
The mixing-based local DA can not only improve model performance in standard supervised learning, but also reduce labeling efforts via self-training~\cite{xiao2022polarmix}. It enables unsupervised domain adaptation without explicit supervision.
Its effectiveness has been studied in text analysis, computer vision~\cite{yuan2021simple, ghiasi2021simple, xiao2022polarmix}, and speech processing~\cite{shorten2021text}, improving the generalizability of models. It synthesizes silver standard labels generated from input data to facilitate learning.
Similarly, in self-supervised learning, DA~\cite{swav2020caron, dino2021caron} even brings more performance improvements than algorithmic techniques like revising model architectures or algorithmic additions~\cite{simclr2020chen, mocov2021chen, byol2020grill, msn2022assran}.
This aspect is crucial for wireless perception because, unlike crawling images and text from the web, collecting sensing data with labels from the real physical world is especially expensive.

\textbf{Wireless sensing} is a promising alternative to cameras for tracking humans, relying on the modalities of radio frequency (RF) signals, such as mmWave and WiFi, or acoustics. The applications range from gesture classification~\cite{hayashi2021radarnet, wang2016interacting}, localization~\cite{zhang2021widar3, qian2018widar2, qian2017widar}, motion detection and pose estimation~\cite{adib2013see, adib20143d, adib2015rf, zhao2018rf, zhao2018through, jiang2020towards}, and fine-grained face reconstruction~\cite{xie2023mm3dface} using RF.
In addition, acoustics detection leverages ubiquitous speakers and microphones with little bandwidth for hand gesture recognition~\cite{amaging2022, wang2020push, gupta2012soundwave, yang2023sequence} and hand tracking~\cite{nandakumar2016fingerio, wang2016device, mao2019rnn, liu2022acoustic, li2020fm}. 

\textbf{Wireless sensing datasets} have become more accessible in recent research. Some focus primarily on a single task, mostly keypoint estimation or action recognition using mmWave~\cite{singh2019radhar, zhao2023cubelearn}. MARS~\cite{marsdataset2021} is one of the pioneers, providing data for rehabilitation using mmWave. mRI~\cite{an2022mri} and MM-Fi~\cite{yang2024mm} are large-scale structures that contain around 160K frames. 
MiliPoint~\cite{cui2024milipoint} first includes all three main tasks in human tracking: user identification, action classification, and keypoint pose estimation, with a total of 49 different actions across 545k frames.
These datasets facilitate research on wireless perception and make DA research possible in this domain by being able to benchmark comprehensively.

\textbf{Data augmentation in wireless sensing} was mentioned in some work, typically tailored to their specific system, yet lacking a generic DA framework with comprehensive evaluations.
For examples of global DA, MiliPoint~\cite{cui2024milipoint} employs a DA, named stack, which involves zero-padding and random resampling. Consequently, the augmented points essentially replicate the original ones with some subset and duplication. This would be a baseline in our experiments. 
\cite{huang2019radiomodule, chen2023radiowavelet, clark2021training, blindsignal2019} use global DA like flipping, rotation, Gaussian noise, or wavelet but only to classify radio signal types, not applicable for sensing.
~\cite{acousticdataset2023} shifts the range profile along the range axis by a small scale, which will make data distorted beyond a certain augmentation scale, leading to decreased performance. 
Besides, \cite{crosssenss2018} uses transfer learning to improve data efficiency across environments but it requires labeled data from the target domain.

Recently, generative models have been explored to synthesize wireless data~\cite{zhao2023nerf2, chen2023rfgenesis}, but are limited to certain tasks or data. \cite{ahuja2021vid2doppler, 2023synthesizemili} simulates RF signals from videos of human actions, but it can only generate scenes from existing video data.
\cite{chen2023rfgenesis} overcomes this issue by using text prompts to synthesize the 3D mesh of humans first, and then simulate mmWave data from the 3D visuals using ray tracing and Diffussion models. Nevertheless, this only supports action-related tasks, because the text prompts can only describe body actions or hand gestures. For example, it can only synthesize data for pose estimation or action recognition tasks, but not for user identification, since it is intractable to accurately depict behavior traits solely by text prompts.
In \cite{zhao2023nerf2}, NeRF is applied to simulate RF signals, but it can only aid simple static tasks of localization or 5G channel estimation rather than motion tracking, since the vanilla NeRF is for static synthesis; and it requires labeled data from the target scene to train the NeRF model.
RF-Diffusion~\cite{chi2024rfdiff} is promising to generalize to tasks but it relies on conditional generation, which may face challenges with longer condition vectors, such as sequences of 90 or more coordinates; the paper only validated on action classification task.
In summary, generative methods require significant effort to train an additional model and struggle to generalize to out-of-distribution data, while mixing-based DA is training-free and excels at enhancing data diversity.

\section{Background: Wireless sensing and data formats}
\textbf{}

\begin{figure*}
    \begin{minipage}{0.64\textwidth}
        \centering
        \includegraphics[width=.75\linewidth]{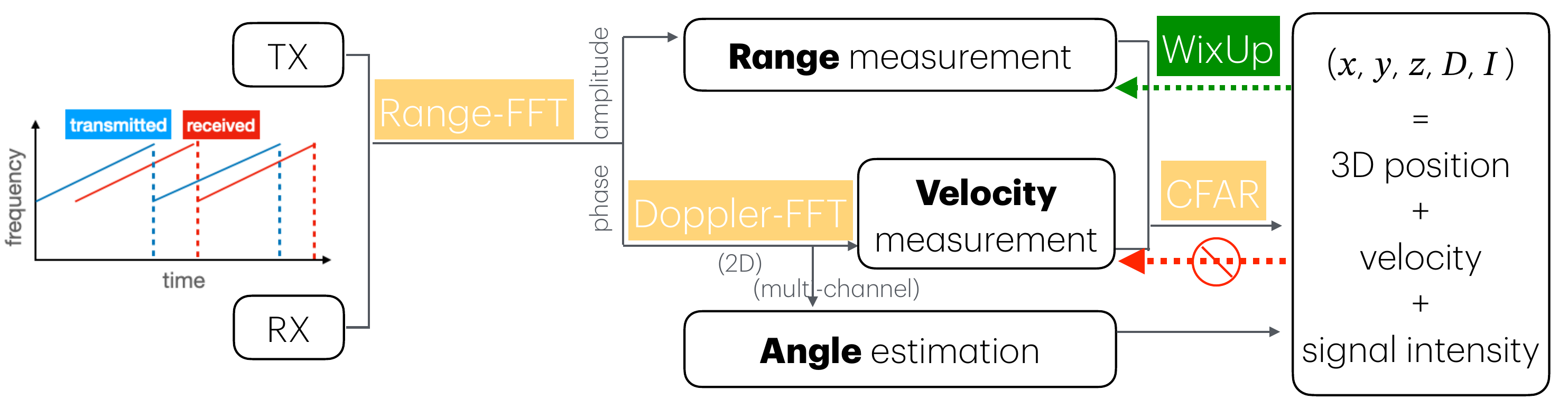}
        \caption{The common data processing pipeline for wireless perception.}
        \label{fig:cfarPipeline}
    \end{minipage}
    \hfill
    \begin{minipage}{0.34\textwidth}
        \centering
        \includegraphics[width=0.8\linewidth]{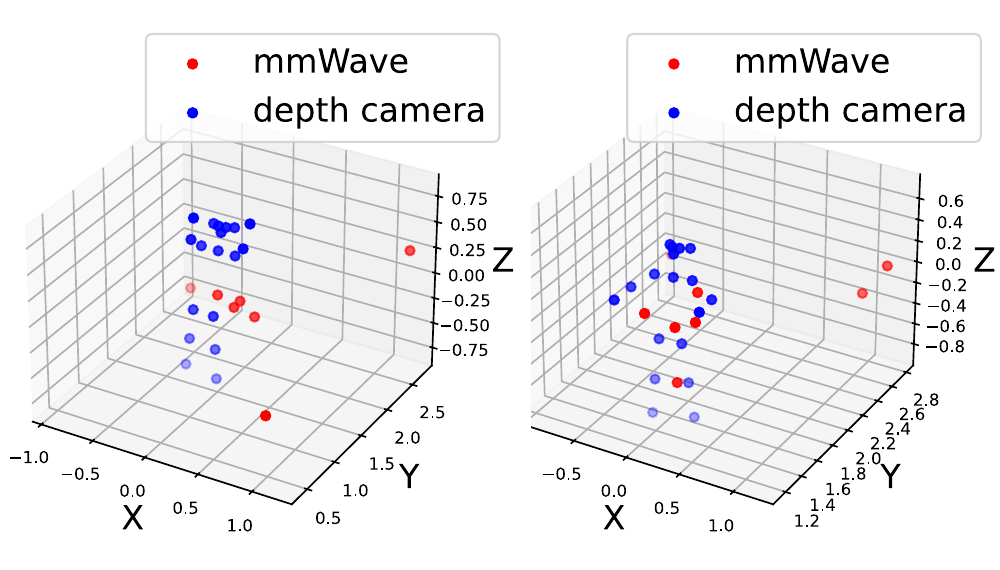}
        \vspace{-4mm}
        \caption{The sparsity issue in wireless.}
        \label{fig:skeleton}
    \end{minipage}
\end{figure*}

In this section, we first provide a brief background about traditional wireless sensing systems to reveal various challenges that lead to \oursys's design.

Most wireless sensing systems, including the standard mmWave radar~\cite{lovescu2020fundamentals}, and many acoustic sensing and WiFi sensing systems use an FMCW radar modulation~\cite{nandakumar2015contactless, li2022lasense, zhao2018rf,zhao2018through} and hence our paper mainly focuses on this pipeline.
Fig.~\ref{fig:cfarPipeline} shows the processing pipeline of FMCW-based wireless sensing systems, which includes de-chirping and extraction of range, angle, velocity, and then 3D Cartesian coordinates.
In detail, first, the transmitter (TX) emits chirps, which reflect off the subject and are then captured by the receiver (RX). These signals are processed through a mixer and filter to produce an Intermediate Frequency (IF) signal (or via algorithms outside the hardware if no mixer is available); A Fast Fourier Transform (FFT) is applied to measure range, followed by Doppler-FFT of the phases along the slow time axis (window axis) to measure velocity. The multi-channel phases can further estimate the angles of arrival. Then a Constant False Alarm Rate (CFAR) filtered the outputs by the noise level. Finally, the ranges and angles can be transformed to Cartesian coordinates along with velocity and signal intensity as 5D times series for downstream applications.
\colorblock{Between modalities, the primary distinction lies in signal frequency. Different modalities’ data exhibits similar characteristics at range profile level.}
This pipeline enables detecting position and movement in human tracking or other applications.

The challenge is the prevalent lack of raw data availability with the majority of public datasets, while the raw data, rich in detail, is critical for performing meaningful and effective DA.
These datasets typically provide only Cartesian coordinates, i.e. the culmination of the data processing sequence.
The issue with Cartesian coordinates is their high-level abstraction, owing to the filtering by CFAR, which leads to a considerable loss of fine-grained information that is imperative for in-depth DA. Moreover, the sparsity issue in the processed point clouds also makes it worse, while the range profile is a relatively dense representation. Unfortunately, it is intractable to reverse the CFAR algorithmically like inverse FFT. 

Hence, our goal is to propose a framework that can transform any data format of wireless signals into range profiles. This is because, unlike Cartesian coordinates, range profile is lossless, and is more interpretable than the raw data, allowing in-depth DA.
The following subsections will detail how we achieve this goal, focusing on developing a way to inverse Cartesian coordinates to a simulated range profile.

\section{System Design}

\begin{figure*}
    \centering
    \includegraphics[width=.67\linewidth]{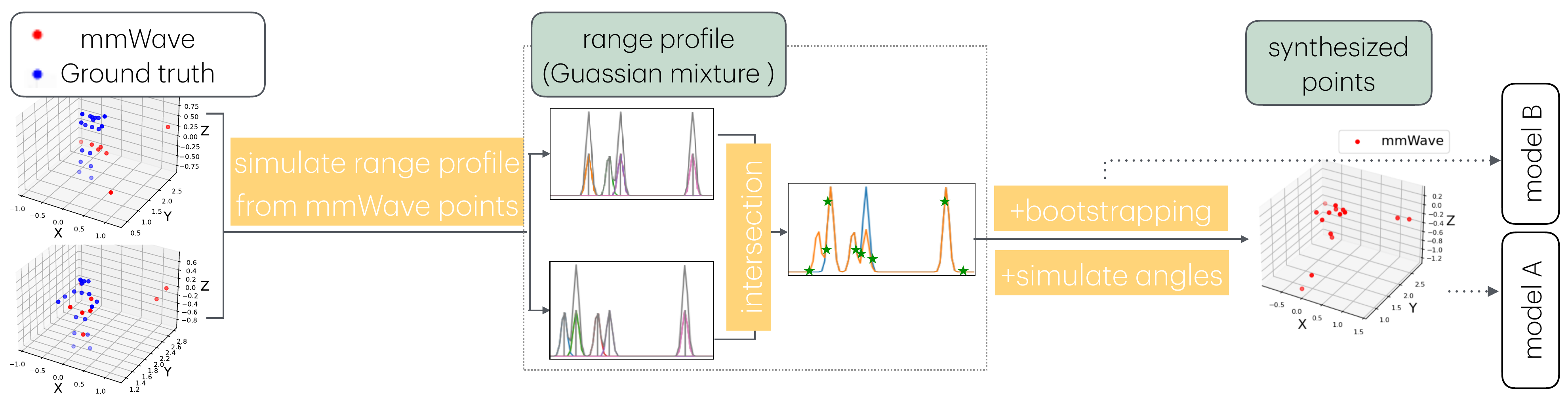}
    \caption{A demonstration of the data mixing pipeline of \oursys.}
    \label{fig:wixupalgo}
\end{figure*}

As illustrated in Fig.~\ref{fig:wixupalgo}, \oursys{} uses a common process of mixing-based DA. So, this section first provides a formal definition of the mix operation in the context of wireless signals, then details the customizations for mixing wireless signals at the range-profile level, and how it is embedded into the common mixing-based process.
In brief, as in Fig~\ref{fig:wixupalgo}, we first describe our solution to transform coordinates into range profiles using Gaussian mixtures, followed by a fault-tolerant mixing method leveraging the intersection of the Gaussians, with only O(n) computation complexity. Bootstrapping can further densify the mixed output. Finally, if the downstream model input is not a range profile but coordinates, we derive them along with the angles based on probability distribution.
Furthermore, to test \oursys's capability of reducing labeling efforts via self-learning, this section also introduces the self-training process that guides our experiment design.

\colorblock{In brief, this pipeline solves the challenges of the heterogeneous data availability of datasets, considering even the irreversible parts. And it is the first mixing DA algorithm customized for wireless signal characteristics.}

\subsection{Mixing-based data augmentation}
\label{sec:mixing-based-da}

A typical mixing-based DA has inputs involving two sets of data: let $m$ and $n$ denote the sensing data and their ground truth. The input for a mixing operation is a pair of data $(m_0, n_0)$ and $(m_1, n_1)$. The output is synthesized data, $(m_0', n_0')$, in the same format as the two input tuples.
Therefore, by iteratively mixing pairs of $(m_i, n_i)$ and $(m_{i+1}, n_{i+1})$ where $i = 0, 1, ...l$ ($l$ is the size of the original data), we could yield $l-1$ of synthesized data by one pass of iteration. Moreover, the distance $d$ between each pair can vary from 1 to 2 or more, i.e. $(m_i, n_i)$ and $(m_{i+d}, n_{i+d})$. Thereby, by iterating with $d=1, 2, ...s$, we scale up the data by $s$ times.

DA focuses on mixing $(m_0, n_0)$ and $(m_1, n_1)$, especially $m_0$ and $m_1$.
Typically, $n$ is usually mixed by a convex combination, i.e. weighted average. In our work, we take the average, which is agnostic to task-specific ground truth format.
For example, in pose estimation, $n$ could be a $19\times3$ matrix, representing 19 body keypoints' coordinates in 3D; thus averaging $n_0$ and $n_1$ equals the 3D skeleton \colorblock{located in the midway between the two original ones.} \colorblocknew{Mixing is joint-wise so that it will mostly result in feasible positions, if the two original positions are not significantly distant within the same time sequence and the frame rate is sufficient. If these preconditions about frame rate and distance are not satisfied, infeasible positions such as shortened or intersected arms might occur. However, supervised learning use cases can mostly meet the conditions. In other cases, solutions such as filtering can be applied to avoid this issue. We will discuss this in detail in Sec.~\ref{sec:discussion} based on our results and theory explanation.}
In tasks like user identification or action recognition, $n$ could be class ID, usually one-hot encoded. So, the average is a probability vector, \colorblock{which has an equal probability of distinct classes, or is identical if classes are the same}.
As for the data $m$, traditional image-based DA uses convex combination for pixel matrices similar to $n$. However, in wireless perception, $m$ could be raw signals, range profiles, 3D point clouds, or 5D time series of $(x, y, z, D, I)$ where $D$ is the Doppler velocity and $I$ is the signal intensity. As discussed above, directly adding them does not meaningfully align with the feature of wireless signals. Instead, \oursys{} aims to transform them all into the range profile level for mixing with lossless information.
Next, we will introduce mixing $m_0$ and $m_1$ via a signal processing pipeline and probability-based algorithms.

\subsection{\oursys{} data mixing pipeline}

\subsubsection{Inverse Cartesian coordinates to simulated range profile using Gaussian mixture}
\textbf{}

Most states in the FMCW data processing pipeline could freely transform to range profile using the standard signal processing algorithms, since they are mostly reversible, such as inverse FFT. However, the CFAR presents a unique challenge as it inherently discards a significant quantity of data, making reversal complex. 
Therefore, we put forth a novel inverse data processing pipeline that operates between the range data and Cartesian coordinates. This enables the application of \oursys{} to any public datasets, irrespective of the data format they are provided in, allowing us to bypass the limitation imposed by the absence of raw data.

The core idea is to simulate the range profile as a Gaussian mixture model of Euclidean distances, as in the left of Fig.~\ref{fig:wixupalgo}. In detail, first, we translate Cartesian coordinates to spherical coordinates as a representation better aligned with the transmission features of wireless signals. Next, dividing the Euclidean range by range resolution yields indices of bins, which are the means in a Gaussian mixture model; the probability density function (PDF) is derived as the range profile, i.e. the statistical distribution of the ranges.
If the hyperparameters of the sensing system have a window size of 512 for de-chirping, and a range resolution of 3.75cm. Then the maximum allowable detection range is 19.2m. Assuming the human joint size is around 10-15cm, we use the standard Gaussian, implying its main lobe falls within the joint size.

\colorblock{Besides, choosing the low-level range profile as the mixing level means that it is also upward compatible with datasets with raw signal available or non-FMCW based algorithms. For instance, other features that can derive from parts of the standard pipeline could also work with our DA, such as Doppler velocity, Synthetic Aperture Radar (SAR) and Channel State Information (CSI). 
This work focuses and further optimizes on human tracking tasks using this standard pipeline, which is currently limited by the availability of open source large wireless datasets for benchmarking.}

\subsubsection{Mixing range profiles using intersection}
\textbf{}

After transforming data into range-profile, whether through the simulation above or calculated from the raw signal using standard de-chirping, we then mix every pair of data into new synthetic data.
We design a mixing strategy by taking the intersection of two range profiles, aligning closely with the physical interpretation of wireless range profiles while also being computationally efficient. 

As illustrated in the center of Fig.~\ref{fig:wixupalgo}, each pair of peaks may intersect midway along the range axis; the height of the intersected side lobe is roughly inversely proportional to the distance between the peaks. This approach involves taking middle points from all possible bisections of two sets of real points, as it is unclear which body joint each real point corresponds to; thus, we cannot restrict middle points to be extracted from identical joints. By considering all possible bisections, we ensure coverage of eligible synthetic points and introduce some "temperature" into the process, borrowed from sampling in VAE. This makes the mixing fault-tolerant.

Instead of employing the closed-form algorithms for the intersection of Gaussian, we utilize an O(n) function to identify intersections. The validity of intersections can be determined using the formula $(b_i-a_i)(b_{i-1}-a_{i-1})<0$, where $a$ and $b$ are the two PDF arrays and $(a_i, a_{i-1})$ denotes a neighbor pair.
This ensures that $ab$ pairs like $(0,0)(0,0)$ and $(0,1)(0,0)$ are excluded. This exclusion is crucial as such pairs do not represent valid intersection points. By implementing this method, we effectively filter out non-intersecting segments by one pass. This streamlines the intersection detection process, enhancing both efficiency and accuracy and does not rely on the complexity of the input Gaussian mixture.

\colorblock{The theory behind this mixing-based DA is the convex combination discussed in ~\cite{zhang2017mixup}. Mixing is directly at the pixel level of images without manually designing features at any higher levels of representation, although it is less comprehensive. Similarly, in the context of wireless data, the physical representations of samples, as well as the environmental factors of sensors such as position and direction, are entangled representations in range profile. In other words, they have unified representations in range regardless of discrepancies in physical settings.}

\subsubsection{Boostrapping high-probability synthesis}
\textbf{}

Besides, we further bootstrap each intersection point weighting on its height. Since the height of the intersection is proportional to its validity as an eligible point, we use it as a weight to resample points around it. This follows the Monte Carlo simulation, where random samples are generated to simulate the properties of a system or to approximate integrals and sums.
Besides, to add more randomness, we also keep the original peaks in the simulated range profile. They are assigned the lowest weight of one and all the weights of the above intersections are elevated by one as well. \colorblock{This step plus the above add-on algorithms in the mixing pipeline helps further improve the fidelity of augmented data. An exemplary fidelity is visualized in Fig.~\ref{fig:wixupalgo}. Owing to the original sparsity nature of wireless signals as shown in Fig.~\ref{fig:skeleton}, visualization can hardly justify the improvements in fidelity. So, we conduct comprehensive benchmark experiments to show its efficacy.}

All aforementioned modules would be ablate studies in our experiment section to justify their effectiveness. For instance, initially we also adopt skewness in Gaussian to represent the multipath induced around the subject, i.e. skewed towards the larger range along the range axis. However, the skewness does not help with improving the model performance, so we discard it from the design of \oursys.

\subsubsection{Enrich range profile with angles based on probability distribution double-paned}
\textbf{}

Finally, if the downstream model input is not a range profile but 3D coordinates, we then need to induce back the coordinates. While, in addition to the ranges, reconstructing coordinates also requires azimuth and elevation angles. Therefore, we use a probability-based method to assign angles to each synthetic range by sampling from a distribution. This distribution is built from the actual angles of the input data pairs.
The probability distribution of angles mirrors the Gaussian mixture used for ranges. Consequently, the final angle at each range is determined by a convex combination of the probabilities associated with major nearby points. For instance, ideally, the midpoint between two ranges will also have averaged angles, making it the midpoint in 3D. This arrangement is in harmony with the expectations set by our mixing method.
Finally, we transform the spherical coordinates to Cartesian as visualized at the end in Fig.~\ref{fig:wixupalgo}.

\subsection{Use \oursys{} for unsupervised domain adaptation via self-training}
\label{sec:uda}


We not only aim to evaluate \oursys{} in standard supervised learning, but also demonstrate its capability for reducing labeling efforts via self-training, achieved through our mixing scheme to generate new data.
Note that non-mixing methods like random scaling are not applicable since they augment one data at a time, unlike mixing two data from different domains to close the domain gap. Mixing-based augmentation enables the creation of more realistic semi-real labels by mixing real data and predicted data, facilitating efficient domain adaptation through self-training.

\textbf{Source domain \& Target domain:} The process is to adapt models trained on a source domain to inference well on a target domain. In this approach, the target domain dataset might have scarce labeled data or even no labeled data.
In step 1 of this approach, the model is first trained with labeled data from the source domain. In step 2, partial target domain data is split out as training data and the rest is for testing. If we make predictions on testing target data using the trained model from step 1, it might likely yield low accuracy. (Note that the labels of testing data are only used for evaluation.) In step 3, the trained model then runs inference on the unlabeled target domain training data, where the output predictions are kept as pseudo-labels for them. Next, in step 4, we take pairs of one source domain data and one target training data and then mix each pair of them as one semi-labeled training data. The mixing algorithm could be customized, such as \oursys{} tailored for wireless sensing data. Finally, in step 5, the new training data further fine-tunes the trained model. Thus, the model performance on the same testing data in the target domain should improve.

\textbf{Use case:} This self-training process proves especially advantageous in wireless sensing, where acquiring new data often demands significant manual efforts, and system performance can significantly drop in new rooms or new users, i.e., unseen domains that are not part of the training data. Currently, to solve this issue, many wireless perception systems need users to undergo calibration/personalization before using it, but obtaining ground truth for the sensing data at the user's end is often unfeasible. Thus, by employing unsupervised domain adaptation through self-training, \oursys{} addresses this practical challenge by enabling calibration/personalization without the necessity of collecting labels from new users or environments.

\section{Experiments}
In this section, we carry out an extensive evaluation of \oursys, divided into three groups of experiments:
\begin{itemize}
    \item We benchmark how \oursys{} enhances model performance in supervised learning across a variety of use cases, no matter what you use for datasets, model architectures, tasks, and sensing modalities. In general, \oursys{} helps increase the size of training data, which constantly provides a significant margin of improvement over no augmentation and outperforms other baseline augmentation methods.
    \item The second group of experiments illustrates \oursys's capability to significantly reduce the need for labeling data by utilizing unsupervised domain adaptation through self-training.
    \item We run ablation studies to verify the effectiveness of the algorithmic modules and hyper-parameters in \oursys{}.
\end{itemize}

The following subsections begin with an elaboration on the experiment setup, followed by results analysis for these three groups of experiments. 

\subsection{Experiment setup}
In the setup, we detail the overall evaluation metrics, task-specific metrics, the baseline methods as a comparison, the datasets we employ, and the implementation we utilize.

\subsubsection{Overall evaluation metrics}
\textbf{}

For both the supervised learning and the self-training experiments, we aim to show that the model could yield a lower error on the validation data after leveraging \oursys, under the same settings including data and training epoch. Moreover, in self-training, \oursys{} should be able to achieve this goal even though some data has no labels.

\textbf{Supervised learning} typically starts with a labeled dataset that is divided into training and evaluation subsets, allowing us to measure performance across both the training and evaluation, using loss or task-specific error metrics.
In our experiments, we ensure only augmenting the training data. More specifically, by augmentation, we can increase the size of the training dataset by two times or more. We then train the model using this enlarged training dataset.
To demonstrate the efficacy of the DA, there should be drops in loss and errors in both the training and evaluation; the training loss should also converge faster than that without the DA. Furthermore, we aim to show that \oursys{} surpasses baseline augmentations by achieving more significant improvements in the evaluation results.

\textbf{Self-training} assumes that not all training and evaluation data have labels; the model trained on limited labeled data might not perform well on the evaluation set. While \oursys{} enables further fine-tuning of the trained model with no-label data; it mixes the no-label data and predicted pseudo-labels with the real data to feed into the model again. Note that the data for mixing is separate from the evaluation set; also, the evaluation set's labels are only for assessing performance, so the labels could be absent in real use cases.
In conclusion, effective self-training with \oursys{} should be able to reduce the error on the evaluation set, in comparison with the model only trained on limited labeled data.

\subsubsection{Task-specific evaluation metrics}
\textbf{ }

To demonstrate the generalizability of our DA across different applications, we selected three common tasks in human tracking: keypoint pose estimation, action recognition, and user identification.
For each task, we train the model with \oursys{} to confirm that it reliably contributes to improvements in the evaluation data, highlighting \oursys' usability in diverse applications.

The evaluation metric for each task is 1) Mean Per Joint Position Error (MPJPE) for keypoint pose estimation and 2) classification accuracy for action recognition or user identification. 
In detail, MPJPE is common for 3D human pose estimation or hand pose estimation, calculating the average Euclidean distance between the predicted and the ground truth positions across key joints. MiliPoint uses mean localization error (MLE), so we also follow this usage in our paper.
Besides, classification accuracy measures the percentage of correct predictions. The goal of action recognition is to classify into one of the predefined actions (e.g., waving hands, jumping up, etc.).
User identification also uses accuracy, whose objective is to recognize a user based on their behavioral or physical traits hidden in the sensor data.

\subsubsection{Baseline augmentation methods}
\textbf{}

We compare \oursys{} with three baselines, no augmentation, conventional global augmentation, and stacking, across all benchmark experiments to demonstrate its effectiveness. \colorblock{Other baselines that come with the original dataset work, such as shifting, are also listed in the following. Besides, temporal signal data augmentation, like stretching, is theoretically equivalent to shifting in the range profile that is included in our paper. Other traditional temporal signal augmentations are not applicable baselines since most of the open-source wireless datasets do not provide full raw data in the time domain. Finally, we also have a section comparing the existing DA methods in the wireless domain, although they are confined to specific tasks, models, or datasets, thus lacking a comprehensive benchmark evaluation.}

\textbf{Baseline Null: no augmentation.} One baseline approach uses solely the original real data without any augmentation.

\textbf{Baseline CGA: conventional global augmentation} Due to the lack of prior generic DA for all tasks and models discussed in this paper, we look into conventional global DA in general 2D and 3D data as our baselines. Therefore, we opt for a conventional global augmentation, named CGA, involving random scaling within the range of 0.8 to 1.2.
Although it can not facilitate unsupervised domain adaptation through self-training due to its non-mixing-based approach, it is applicable in all the benchmarking experiments we will delve into in the following. 
Specifically, in pose estimation, scaling is applied to both sensing point clouds and ground truth points, while for user and action recognition, scaling is not applied to class labels.

\textbf{Baseline Stack: random duplication.} The MiliPoint~\cite{cui2024milipoint} dataset also introduces a simple DA technique, named stack. This method involves zero-padding each frame to standardize the number of points. Subsequently, they randomly resample from these points, serving as one data duplication. According to the original paper, this procedure is repeated several times for each frame: five times for tasks involving pose estimation and action recognition, and fifty times for the user identification task.
Intuitively, this approach not only ensures a uniform input size but also enhances the data diversity through randomness. However, our reproduction results show that it actually negatively impacts the model performance in many cases. So, we only run it in initial benchmarking experiments and refrain from the rest.

\textbf{\oursys$^{+}$: a combination of \oursys{} and CGA.} Except for the self-training experiments, it is valid to use \oursys{} on top of Baseline CGA in benchmarking experiments. In practice, it is common to employ multiple DA methods simultaneously in many machine-learning applications. Therefore, we combine \oursys{} with CGA as another comparison with only \oursys.
However, in our experiment results, we observed that the inclusion of CGA, noted as \oursys$^{+}$, does not always bring additional performance enhancements, a point we will delve into in the forthcoming analysis section. Overall, \oursys{} itself constantly increases performance beyond the capabilities of individual baselines in most cases. 


\subsubsection{Datasets}
\label{sec: datasets}
\textbf{ }

In our experiments, we leverage three mmWave datasets and one additional acoustic dataset for the cross-modality experiment. Specifically, they are three publicly available mmWave human tracking datasets alongside an acoustic hand-tracking dataset collected from our prior research. This diverse selection could demonstrate the generalizability of \oursys{} across datasets and sensing modalities.

\textbf{MiliPoint Dataset}~\cite{cui2024milipoint} focuses on low-intensity cardio-burning fitness movements.
With 49 distinct actions and a massive dataset of 545,000 frames from 11 subjects, it surpasses previous datasets in both action diversity and data volume. It also covers pose estimation and user identification. We use this dataset as the primary test bed for our benchmarking experiments.
The data collection process considers factors such as movement intensity and diversity. Participants perform a series of movements while monitored by a mmWave radar and a Zed 2 Stereo Camera for the ground truth, \colorblock{at a frame rate of around 24Hz}. The Texas Instrument IWR1843 mmWave radar, a common choice for wireless perception research, operates between 77 GHz to 81 GHz with a chirp duration of 100 us and a slope of 40 MHz/us, with a large bandwidth of 4GHz for a 4cm range resolution.
The Stereo Camera provides ground truth of 18 3D keypoints, offering reliable accuracy for their settings at both 3 and 15 meters away.
However, this dataset consists solely of 3D point clouds without the raw mmWave data, leaving around 8-22 points per frame. So, we could show \oursys{} can inversely use the processed data for DA.

\textbf{MARS Dataset} (Millimeter-wave Assistive Rehabilitation System)~\cite{marsdataset2021} is a pioneer work providing large-scale datasets in wireless sensing, designed for the rehabilitation of motor disorders utilizing mmWave.
This work comes with a first-of-its-kind dataset of mmWave point cloud data, featuring 70 minutes of 10 different rehabilitation movements performed by 4 human subjects, providing 19 human keypoints and 40,083 labeled frames alongside video demonstrations made public.
One common Texas Instrument IWR1443 Boost mmWave radar runs data acquisition at 76-81GHz. The chirp duration is 32us with a slope of 100MHz/us, facilitating a range resolution of 4.69cm and a maximum detection range of 3.37cm.
Additionally, a Kinect V2 with infrared depth cameras captures ground truth, collocated with the radar and synchronized with the Kinect’s fixed sampling rate of 30 Hz.
Similarly, their signal processing only keeps the first 64 points, consisting of a 5D time-series point. We take the subset of solely the coordinates to run the same experiment set up with the other mmWave dataset we use.

\textbf{MMFi Dataset} stands out as another large-scale wireless dataset~\cite{yang2024mm}, featuring the first multi-modal non-intrusive 4D human dataset, including mmWave radar, LiDAR, WiFi CSI, infrared cameras, and RGB cameras, for the fusion of sensors and multi-modal perception.
The total has 320,000 synchronized frames across five modalities from 40 human subjects with 27 categories of daily and rehabilitation actions, providing a valuable view for both everyday and clinical research in human motion.
The radar is Texas Instrument IWR6843 60-64GHz mmWave, whose detailed parameters were not disclosed. \colorblock{The RGB and depth cameras both have a frame rate of 30Hz and the radar is set correspondingly.} Moreover, a novel mobile mini-PC allows data collection in diverse environments. So, we use this dataset in our self-training experiments to test unsupervised domain adaptation across environments.

\textbf{Acoustic Dataset: }
We collected an acoustic dataset in ~\cite{acousticdataset2023}, which is a sensing platform for human hand tracking, employing the same FMCW signal algorithms of mmWave radars. It features fine-grained continuous tracking of 21 highly self-occluded finger joints in 3D.
Data was collected from 11 participants across three environments, totaling 64 minutes of selected hand motions, covering expressive finger joint movements.
The hardware comprises a development microphone array board, a speaker, and a Leap Motion infrared camera for ground truth (\colorblock{frame rate is 90$\sim$110Hz}). The 7-channel mic array shares the same layout and sensitivity specifications as Amazon Echo 2. The system emits ultrasound modulated into 17k-20kHz chirps with a 10ms duration, achieving a range resolution of 3.57mm, owing to the low speed of sound and the custom de-chirping.
This dataset helps demonstrate \oursys's versatility across multiple wireless modalities and flexibility in handling other input data formats.

\colorblock{Note that the wireless dataset’s frame rate is usually limited to its ground truth camera’s frame rate, which is typically around 30Hz. As the chirp duration reported above, the chirp rate with no idle time could be up to a few KHz for mmWave sensing, and 100Hz for acoustic sensing. While the wireless signal data could have a higher frame rate, the experiment settings are aligned at a similar scale with ground truth camera’s frame rate. After all, one advantage of our method is that it is designed to work in the absence of this kind of parameter.}

\subsubsection{Implementation}
\textbf{}

To run a wide range of experiments across datasets, tasks, model architectures, and baselines, we need a flexible code framework. Since, two of the three mmWave datasets, MARS and MMFi, do not come with a full code base, we rewrite the codebase of MiliPoint as a uniform framework for testing all datasets. It supports a variety of model architectures including DGCNN, PointNet++, and PointTransformer, along with three tasks.\colorblock{We published our code  of WixUp framework on Github\footnote{https://github.com/lydhr/WixUp}}, aiming to facilitate further research in DA within this domain.

For context, all experiments are executed on a GPU server equipped with NVIDIA Quadro RTX 8000. 
Wherever possible, we adhere to the hyper-parameters outlined in the original dataset papers to ensure accurate and equitable reproduction. It covers the hyper-parameters such as the learning rate for training, the batch size, pre-processing schemes, and even the random seed, as well as the evaluation metrics. Except for one point, we disable the random shuffling of data before splitting into training and testing sets, which does not align with the reality that the testing data only happens after the time of training data, although no-shuffle might decrease our accuracy.
Most experiments utilize subsets of the original data to facilitate extensive ablation studies and benchmarking. We ensure that any comparisons are made align with the reproduced original settings under our framework.

\begin{table}[]
\begin{tabular}{lccc}
\hline
 & \multicolumn{1}{l}{\begin{tabular}[c]{@{}l@{}}Identification\\ (Acc\%)\end{tabular}} & \multicolumn{1}{l}{\begin{tabular}[c]{@{}l@{}}Action\\ (Acc\%)\end{tabular}} & \multicolumn{1}{l}{\begin{tabular}[c]{@{}l@{}}Keypoint\\ (MLE in cm)\end{tabular}} \\ 
 \hline
Report & 77.65 & 13.61 & 16.53 \\
Reproduction & 80.90 & 15.88 & 16.93 \\ \hline
\end{tabular}
\caption{Our code base implementation can reproduce the results reported in the MiliPoint~\cite{cui2024milipoint} dataset.}
\label{tab:reproductionExp}
\end{table}

Table~\ref{tab:reproductionExp} illustrates our reproduction result of MiliPoint, wherein we run 50 epochs on the full data set for each of the three tasks using the DGCNN model. It runs less than 10 hours on a single RTX 8000 for each task.
Our reproduction yields slightly better results for user identification and action recognition in terms of top-one classification accuracy, and it maintains a similar MLE for 9-point keypoint pose estimation. In conclusion, the reproduction result is verified to align with the reported results in the original paper.
Besides, the errors here are usually higher than those in the following benchmarking because those are trained with a subset of data and five times fewer training epochs, to facilitate the extensive benchmarking experiments of various scenarios.

\subsection{Benchmark the generalizability of \oursys{} in supervised learning}
\textbf{ }

Following the above setup, first, we assess how \oursys{} boosts model performance in supervised learning across diverse scenarios, encompassing datasets, model architectures, tasks, and sensing modalities. Broadly, \oursys{} expands the training data size for supervised learning, consistently delivering notable enhancements over the no-augmentation result, surpassing other baseline augmentations.

\subsubsection{Generalize across datasets}
\textbf{ }

The experimental results presented in Table.~\ref{tab:cross-dataset} showcase the performance across three distinct datasets: MiliPoint, MARS, and MMFi, as we elaborated above in the experiment setup section~\ref{sec: datasets}.
To make the training and evaluation data size equitable across datasets, we task a 20\% subset of MiliPoint, and 20\% of MMFi; then they all have around 40k data in total split into 80\% for training and 20\% for testing.
The numbers are errors of MLE in cm for keypoint pose estimation, trained with the DGCNN model.

The initial observation from the results reveals that, as indicated by the percentages in parenthesis, all augmentation methods could outperform the Baseline Null, which has no augmentation. For example, (+26.95\%) means the \oursys$^{+}${} reduces the error from 28.89 to 21.10 by 26.95\% percent. Except for Baseline Stack, proposed by the original MiliPoint paper, which actually negatively impacts accuracy. 
Consequently, we refrain from running this baseline in subsequent benchmarking efforts.
Secondly, we observe that \oursys{} consistently delivers greater improvements compared to Baseline CGA. Furthermore, the additional integration of \oursys{} and CGA, noted as \oursys$^{+}${}, leads to further enhancements in most scenarios. However, as previously mentioned, the addition of CGA does not always yield improvement, potentially due to the unstable nature of random scaling in CGA, a topic we will delve into in subsequent results' discussions.
In summary, our system consistently delivers the most significant performance improvements across the three benchmarking datasets. This underscores its robustness and ability to generalize effectively across diverse datasets.

\setlength{\tabcolsep}{1.6pt}

\begin{table}[]
\begin{tabular}{llll}
\hline
 & MiliPoint & MARs & MMFi \\ \hline
Null & 24.36 & 28.89 & 28.53 \\
Stack & 25.16(-3.25\%) & 28.89(-0.02\%) & 28.53(+0.01\%) \\
CGA & 23.26(+4.55\%) & 22.89(+20.76\%) & 26.25(+8.00\%) \\
\oursys & 23.25(+4.58\%) & 22.65(+21.60\%) & 25.84(+9.42\%) \\
\oursys$^{+}$ & \textbf{23.10 (+5.18\%)} & \textbf{21.10(+26.95\%)} & \textbf{25.60(+10.28\%)} \\

\hline
\end{tabular}
\caption{\oursys{} effectively reduces the error across diverse datasets and surpasses other baselines.}
\label{tab:cross-dataset}
\end{table}

\setlength{\tabcolsep}{3pt}

\subsubsection{Generalize across model architectures}
\textbf{ }

\begin{table*}
\centering
\begin{tabular}{llllllll}
\hline
 & \multicolumn{3}{c}{DGCNN} &  & \multicolumn{3}{c}{PointTransformer} \\ \cline{1-4} \cline{6-8} 
 & \multicolumn{1}{c}{MiliPoint} & \multicolumn{1}{c}{MARS} & \multicolumn{1}{c}{MMFi} &  & \multicolumn{1}{c}{MiliPoint} & \multicolumn{1}{c}{MARS} & \multicolumn{1}{c}{MMFi} \\ \cline{1-4} \cline{6-8}
Null & 24.36 & 28.89 & 28.53 & & {17.15} & {25.64} & {24.69} \\ 
CGA & 23.26(+4.55\%) & 22.89(+20.76\%) & 26.25(+8.00\%) & & 16.88(+1.60\%) & 20.74(+19.10\%) & 24.03(+2.68\%) \\
\oursys & 23.25(+4.58\%) & 22.65(+21.60\%) & 25.84(+9.42\%) & & 16.81(+1.97\%) & \textbf{20.35(+20.64\%)} & \textbf{22.73(+7.96\%)} \\
\oursys$^{+}$ & \textbf{23.10(+5.18\%)} & \textbf{21.10(+26.95\%)} & \textbf{25.60(+10.28\%)} & & \textbf{16.67(+2.79\%)} & 21.35(+16.72\%) & 23.26(+5.80\%) \\
\hline
\end{tabular}
\caption{\oursys{} consistently improves performance across model DGCNN and PointTransformer, even outperforms CGA+\oursys{} sometimes owing to the instability of CGA.}
\label{tab:cross-models}
\end{table*}

The experimental results in Table.~\ref{tab:cross-models} highlight the performance across two distinct model architectures: DGCNN and Pointformer, tested on all three datasets of MiliPoint, MARS, and MMFi. The numbers in the table are MLE errors in cm for keypoint pose estimation.
In conclusion, \oursys{} consistently yields more improvements compared to Baseline CGA.
Moreover, the supplementary incorporation of \oursys{} alongside CGA (\oursys$^{+}${}) brings additional enhancements across most scenarios, except for PointTransformer on MARS and MMFi where \oursys{} along performs better. The incorporation of CGA, we think, doesn't consistently result in improvements, possibly owing to the unstable nature of random scaling within CGA. The instability arises from the significant randomness inherent in random scaling, despite its straightforward usage in this context.

To elaborate, the chosen model architectures collectively represent diverse approaches and have greatly influenced advancements in point cloud analysis. DGCNN captures local and global features through graph convolutions, Pointformer utilizes self-attention mechanisms inspired by transformers.
Alongside DGCNN and Pointformer, we actually also tested with PointNet++ model. However, we encountered difficulties reproducing the results reported in the original paper; our errors were significantly larger. As a result, we refrain from depicting it here as a valid setting to test \oursys, although \oursys{} outperforms other baselines anyway, regardless that their overall errors are high.

In summary, \oursys{} is robust and may outperform its combination of CGA (\oursys$^{+}$) due to its stability. Overall, this table of results highlights its flexibility as a general module, applicable to downstream tasks regardless of their methods.

\subsubsection{Generalize across human tracking tasks}
\textbf{ }

\begin{figure*}
    \begin{minipage}{0.54\textwidth}
    \centering
        \begin{tabular}{llll}
        \hline
          &\multicolumn{1}{l}{\begin{tabular}[c]{@{}l@{}}Keypoint\\ (MLE in cm)\end{tabular}} &  
          \multicolumn{1}{l}{\begin{tabular}[c]{@{}l@{}}Identification\\ (Acc\%)\end{tabular}} & 
          \multicolumn{1}{l}{\begin{tabular}[c]{@{}l@{}}Action\\ (Acc\%)\end{tabular}} \\
        \hline
        Null & {24.36} & 0.8766 & 0.1305 \\
        CGA & 23.26(+4.55\%) & 0.9097(+3.77\%) & 0.2143(+64.17\%) \\
        \oursys & 23.25(+4.58\%) & 0.9031(+3.03\%) & 0.1924(+47.44\%) \\
        \oursys$^{+}$ & \textbf{23.10(+5.18\%)} & \textbf{0.9229(+5.29\%)} & \textbf{0.2405(+84.25\%)} \\
        \hline
        \end{tabular}
        \captionof{table}{\oursys{} generalize well across three tasks.}
        \label{tab:cross-tasks}
    \end{minipage}
    \hfill
    \begin{minipage}{0.45\textwidth}
    \centering
        \begin{subfigure}[t]{.45\linewidth}
        \includegraphics[width=\linewidth]{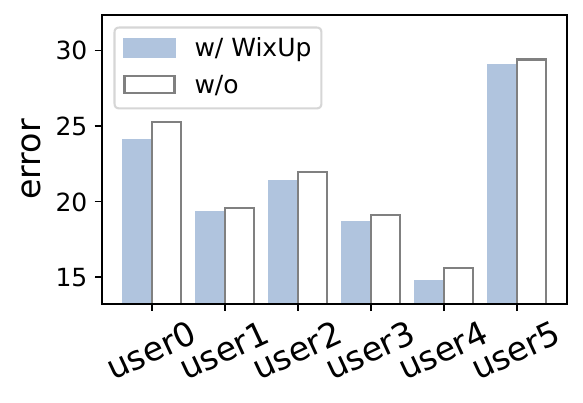}
        \end{subfigure}
        \begin{subfigure}[t]{.45\linewidth}
            \includegraphics[width=\linewidth]{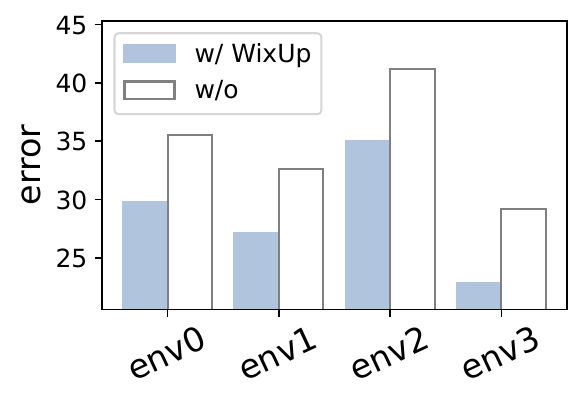}
            \label{fig:selftrain-env}
        \end{subfigure}
        \vspace{-3mm}
        \captionof{figure}{Self-training results for unsupervised domain adaptation across users and environments.}
        \label{fig:selftrain}
    \end{minipage}
\end{figure*}

Table.~\ref{tab:cross-tasks} illustrates the performance across three human tracking tasks: keypoint pose estimation, action recognition, and user identification. The numbers in the table are task-specific errors. Within the subset of MiliPoint used in benchmarking, it has two unique users and 49 unique actions.

In brief, \oursys{} outperforms Baseline CGA in most cases. Furthermore, \oursys$^{+}${}, the combination of \oursys{} with CGA, leads to further improvements. Notably, while action recognition with \oursys{} falls short of CGA, \oursys$^{+}${} surpasses CGA by a significant margin, reaching 84\% of improvement. The reason could be that the overall accuracy in action recognition is low, so the percentage might diverge dramatically. \colorblock{Moreover, expanding the augment data size by tweaking \oursys{} brings further drops in errors, which we will ablate study in the following section. Besides, \oursys{} is a mixing-based local augmentation, providing more flexibility than global augmentation like CGA.}

The choice of keypoint pose estimation, action recognition, and user identification tasks for evaluation in wireless perception underscores their fundamental relevance in various applications. They play a crucial role in scenarios such as security surveillance and healthcare monitoring. The inclusion of these tasks in the evaluation demonstrates the potential of \oursys{} for widespread deployment in wireless perception applications.

\subsubsection{Generalize across sensing modality\&data format}
\textbf{ }

To assess \oursys{} with different wireless sensing modalities and raw data formats, we test with this acoustic dataset along with its dedicated code base using a CNN+LSTM model. As outlined in ~\ref{sec: datasets}, this dataset entails an acoustic sensing system designed for tracking 21 finger joints in 3D. 
To apply \oursys{} with acoustic range profiles, we simply bypass the step of simulating from coordinates to range profiles. Instead, we directly intersect the range profile to generate new range profiles as model input. In contrast, the Baseline CGA does not directly apply to the range profile anymore. So, we keep the straightforward DA method by slightly shifting the range profile along the range axis; and we use the original accuracy reported in the paper as a baseline comparison with \oursys.

In summary, the original mean absolute error (MAE) is 13.93mm for user-dependent pose estimation. By running the same test via \oursys{}, we achieved a better MAE of 10.56mm(+24.19\%), which closely approaches the best results reported by the paper achieved by user-adaptive testing.
\colorblock{In this regard, we observe that generalizing across modalities should be less challenging than that across data formats. Theoretically, this could be attributed to the fact that the primary distinction between modalities lies in signal frequency. When processed by the standard pipeline in Fig.~\ref{fig:cfarPipeline}, different modalities’ data exhibits similar characteristics at the range profile level, which are generally a mixture of Gaussians as visualized in Fig.~\ref{fig:wixupalgo}.}

\subsection{\oursys{} reduces labeling efforts by unsupervised domain adaptation via self-training}
\textbf{ }

Beyond testing \oursys{} in standard supervised learning, we extend our objective to show its capability of self-training, founded on its scheme of mixing data to synthesize new data. In this subsection, another group of experiments illustrates \oursys's capability to significantly reduce the effort for labeling data by utilizing unsupervised domain adaptation (UDA) through self-training.

Moreover, it is worth mentioning that non-mix-up methods like CGA and stacking can NOT be applied to UDA for self-training. Because they augment one data point at a time instead of mixing two data points from two distributions. In contrast, synthesizing \colorblock{semi-real labeled data} by mixing two points allows for the creation of more diverse and realistic samples, aiding in domain adaptation via self-training. 

In practice, this self-training process is particularly advantageous in wireless sensing, where acquiring new data often demands significant manual efforts, and system performance might drop in new users or environments.

\colorblock{To clarify, domain adaptation is a broad topic encompassing a wide range of approaches beyond data augmentation, such as adversarial training, contrastive learning, feature transformation, etc. which are beyond the scope of this paper. This work primarily focuses on proposing a data augmentation framework for wireless sensing, with domain adaptation being one of its usages.}

\subsubsection{Unsupervised domain adaptation across users}
\textbf{}

In the context of user domain adaptation, the steps of UDA via self-training are the same as the above~\ref{sec:uda}. 
1) The model undergoes training using labeled data sourced from the training users, constituting the source domain. 2) The device is sold to a new user, who starts by inputting some new data designated for self-training. 3) The trained model performs inference on the unlabeled data from the new user, generating output predictions that are retained as pseudo-labels. 4) Pair one sample from the source domain data with one from the new user data, and mix them via \oursys{} to form semi-labeled training data. 5) The newly created training data is utilized to fine-tune the trained model. Consequently, the model's performance on the future new user data is expected to exhibit improvement.
In other words, the training and labeling are confined to the efforts of the sensor developer; when a user buys a new device, they can effortlessly improve the device performance without labels of their data, since the device might not equipped with ground truth cameras.

To prove this, we train and evaluate UDA across users in MiliPoint by leave-one-user-out at a time. The left in Fig.~\ref{fig:selftrain} depicts the error of MLE in cm for keypoint pose estimation for six separate users. While some users get great model performance overall, such as user 4. We do see constant performance improvement with \oursys{} than that without (w/o) \oursys{}. The average improvements are 3.04\%, ranging from 1.26\% to 5.83\%.

\subsubsection{Unsupervised domain adaptation across environment}
\textbf{}

Beyond the need for new users, new environments are often another essential factor that impacts the sensing system's performance. New environments here refer to new rooms, or the same room with furniture or nearby metal objects rearranged. The multi-path reflections from the environment might influence the received sensing signals, which could distort learning-based sensing algorithms.
In the context of environment domain adaptation, the steps of UDA via self-training are similar to the cross-user process. In short, when employing the device in a new room, users can effortlessly improve the device's performance without any labeling.

To validate this, we use the MMFi dataset, which has four clearly labeled scenes in data collection. Note that new scenes also imply new users. Therefore, we train and evaluate UDA across scenes in a leave-one-scene-out manner. The right in Fig.~\ref{fig:selftrain} depicts the error of MLE in cm for keypoint pose estimation. All four scenes yield a large margin of performance improvement with (w/) \oursys{} than that without (w/o) \oursys. The average improvements are  17.45\%, ranging from 15.04\% to 21.73\%.

In summary, this mix-up-based approach for UDA in self-training has been empirically shown to lead to better performance in closing the domain gap and improving model generalization across users and environments. In future work, if we have one source dataset sharing the same label format as a target dataset, we could even self-train across datasets or even modalities, without having labels for the target dataset.

\subsubsection{Comparison with generative-model-based DA}
\textbf{}

Table~\ref{tab:generative-compare} summarizes a comprehensive comparison with generative-model-based DA by reporting their experiment results and applicable tasks, showing absolute improvements in percentages and relative improvements (in parentheses). NA means the model is not applicable to that task; dash - means theoretically applicable but not reported in their experiments. To make the comparison relatively fair despite differences in datasets or tasks, we report a range of all results in this scenario.

\colorblock{All these works test on their own collected dataset, focusing on RF signals. Each of them is confined to certain tasks and dataset, making them invalid for a comprehensive comparison to our benchmark results as baselines.}
Most methods could facilitate cross-domain adaptation. However, RF-Diffusion~\cite{chi2024rfdiff} was only tested on action classification tasks, as it relies on conditional generation, which may face challenges with longer condition vectors, such as sequences of 99 coordinates, which have yet to be validated. RF-Gen~\cite{chen2023rfgenesis} demonstrated strong improvements in domain adaptation but cannot be applied to identification tasks, as its text prompt cannot depict this complex feature. NeRF2~\cite{zhao2023nerf2}, focused on static scene reconstruction, has no experiments in dynamic or cross-domain scenarios where its method would significantly degrade. \colorblock{In general, these generative methods are limited to certain tasks or data, whose experiments are on their own collected data and compare with one no-synthesis baseline, or up to three other generative models. In contrast,} \oursys{} demonstrates strong generalization to various tasks, \colorblock{along with validation across a variety of datasets, model architectures, modalities, and baseline comparisons as above}; moreover, our domain adaptation can even be achieved in an unsupervised manner through self-training, which produces higher-quality semi-labeled synthetic samples.

\begin{table*}[]
\begin{tabular}{llll}
\hline
 & \textbf{Action (body)} & \textbf{Identification} & \textbf{\begin{tabular}[c]{@{}l@{}}Domain adaptation\\ (Action/Keypoint)\end{tabular}} \\ \hline
\textbf{NeRF2~\cite{zhao2023nerf2}} & NA & NA & NA \\
\textbf{RF-Gen~\cite{chen2023rfgenesis}} & - & NA & (6.04\%$\sim$56.42\%) \\
\textbf{RF-Diffusion~\cite{chi2024rfdiff}} & 1.8\%$\sim$8.7\%(1.98\%$\sim$11.01\%) & - & (5.34\%$\sim$16.20\%) \\
\textbf{WixUp} & 11\%(84.25\%) & 2.0\% (5.29\%) & (1.26\%$\sim$21.73\%) \\ \hline
\end{tabular}
\caption{Comprehensive comparison with generative DA. [NA = not applicable; - = not reported]}
\label{tab:generative-compare}
\end{table*}

\subsection{Ablation study of \oursys{}}
\textbf{ }

Finally, this subsection shows the experiments of ablation study over \oursys's algorithmic modules and hyper-parameters to validate their effectiveness. 

\begin{figure*}
    \begin{minipage}{0.43\textwidth}
        \centering
        \begin{subfigure}[t]{.46\linewidth}
            \includegraphics[width=\linewidth]{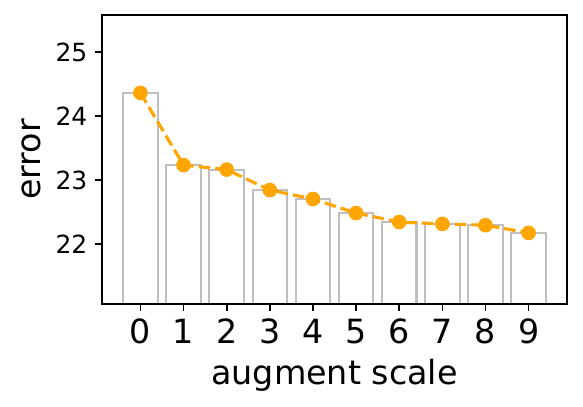}
            \end{subfigure}
        \begin{subfigure}[t]{.275\linewidth}
            \includegraphics[width=\linewidth]{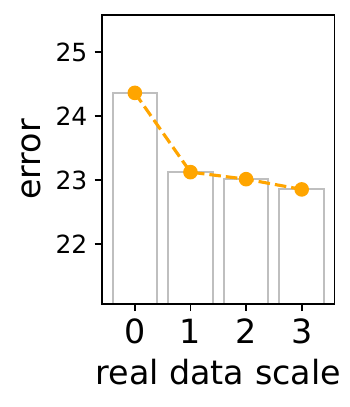}
            \end{subfigure}
        \vspace{-10px}
        \captionof{figure}{Expanding the augment data size by tweaking \oursys{} brings further drops in errors.}
        \label{fig:ablate-scale}
        
    \end{minipage}
    \hfill
    \begin{minipage}{0.55\textwidth}
        \centering
        \begin{tabular}{llll}
        \hline
         & MiliPoint & MARS & MMFi \\ \hline
        Null & 24.36 & 28.89 & 28.53 \\
        vanilla & 24.29(+0.29\%) & 22.8(+21.08\%) & 25.93(+9.11\%) \\
        +Boostrap & 23.25(+4.58\%) & 22.65(+21.60\%) & 25.84(+9.42\%) \\
        +CGA & \textbf{23.10 (+5.18\%)} & \textbf{21.10(+26.95\%)} & \textbf{25.60(+10.28\%)} \\
        \hline
        \end{tabular}
        \captionof{table}{Ablation study on incremental versions of \oursys{} proves the effectiveness of each component.}
        \label{tab:ablate-components}
    \end{minipage}
    
\end{figure*}

\subsubsection{Augmentation size}
\textbf{}

First, we investigate the impact of mixing distance, which refers to the number of interval frames between the two samples to be mixed. For example, distance=2 means each pair of mixing data is sampled from one real data along with its neighbor located two frames ahead.
\colorblock{When increasing the maximum allowable distance, data augmentation scale also increases. 
Scale refers to the factor by which the augmentation data size exceeds the original data size. For instance, scale=6 means we mix pairs whose distance=1,2,3,4,5,or 6; the mixing at each distance yields new data equivalent in size to the original data (ignoring the trivial number of unreachable pairs at the tail, around only distance-1 pairs), thus scale=6 implies that the augmentation data is 6 times the size of the original.}
Usually, stacking pairs from varying distances results in a larger augmentation size and thus greater diversity in data distribution. 
Although augmenting more data typically improves accuracy, there may be a turning point or plateau point where improvements slow down.
As shown in Fig.~\ref{fig:ablate-scale}, we increase the distance from one to ten, ~\colorblock{that is equivalent to a maximum distance of around 0.3s.} It is in the context of key point estimation in MiliPoint with the error as MLE in cm. Excessive augmentation makes the benefits slow down \colorblock{when scale>6} but, impressively, it is still decreasing; The slow down is possibly due to the decreased accuracy in mixing distant pairs. It could also be because the benefits of enriching data distribution reach a limit, ceasing to other primary bottlenecks such as learning algorithms' limited scaling law.
To clarify, the above experiments only augment by a distance of one by default, in order to facilitate the experiments; thereby, we could expect more drops in their error when excessively fine-tuning a single job.
Besides, since we used a 20\% subset of real data for the ablation experiments, we demonstrate that increasing real data (up to three additional scales, excluding the test data) reduces error similarly. This shows our augmentation data has equivalent realness whereas augmentation can be scaled up to three times or more, unlimited to the amount of real data available.
\colorblock{In general, to select an optimal augmentation, developers should consider the dataset frame rate, downstream model and task, as well as the computation resource constraints.}

\subsubsection{Effectiveness of algorithmic components}
\textbf{ }

We also perform several ablation studies to examine the contribution of the components in the proposed \oursys. 
To recap, 1) the vanilla version of \oursys{} conducts an intersection operation on two range profiles. It first simulates each range profile from coordinates as Gaussian mixture and then inversely maps them back with probability-based angles. 2) Secondly, to further increase the number of generated points from the intersection step, we randomly sample around the intersections based on the quality of the intersection as weights and also sample the original points with minimal weight. This process is referred to as bootstrapping. 3) Finally, in benchmark experiments, we utilize CGA (random scaling) as a baseline and add it on top of \oursys{}, denoted as \oursys$^{+}${}.
As shown in Table.~\ref{tab:ablate-components}, each row represents the incremental adoption of these three versions of \oursys{}, along with the Baseline Null as a comparison. The incremental drops of errors in rows demonstrate that each component contributes slightly to the enhancement of model accuracy, thereby validating the effectiveness of incorporating these components.
Besides, in the early stages of our research, we proposed other components such as skewed Gaussian, which theoretically seemed promising but did not help with improving the results. Consequently, they trust our trial and error ablation study results and decided not to integrate them into the final version of \oursys{} presented in this study.

\section{Discussion}
\label{sec:discussion}

In this paper, we focus on human tracking tasks within wireless perception, since its popularity brings ample open-source large datasets and methods, ready for us to conduct comprehensive DA research. However, it's important to acknowledge that wireless perception encompasses a broader spectrum of applications beyond human tracking, for instance, healthcare monitoring, smart agriculture, and supply chain management. We hope to extend \oursys{} to these applications once they mature with more resources in the future.

Moreover, WiFi is also a promising modality for wireless applications, many of which adhere to similar FMCW-based modulation and processing pipelines. However, the costly hardware and excessive installation efforts make this area have less open-source data in public. For instance, \cite{jiang2020towards, zhao2018rf,zhao2018through} are notable works on this topic but only published the code. We hope \oursys{} can embark on open research in this modality as well.

Besides FMCW, other types of signals exist, such as those involving sine wave signals with their phase or Doppler characteristics. Since we emphasize generalizability in this work, we surveyed wireless hardware user manuals and related work and then chose to focus on the current settings; this ensures our DA widely covers industrial and research wireless sensing systems.
We encourage other types of signals to develop their custom DA on top of \oursys{} such as customizing their transforming to range profiles. 

In our exploration of self-training, we conducted experiments involving cross-user and cross-environment scenarios using the same dataset. However, self-training can also be powerful when mixing data from multiple datasets. However, the lack of public information also hinders us from investigating it in this work; specifically, the disclosure of ground truth format such as joint order is not always detailed. Nevertheless, it is a compelling direction for future research. 
Note that cross-dataset differences, like hardware variations or signal parameters, do not impede cross-dataset self-training in our approach, owing to its ability to simulate a unified format for sensing data. 

\colorblocknew{As mentioned in Sec.~\ref{sec:mixing-based-da}, our approach involves mixing the joint-wise positions in the midway. This could produce infeasible joint positions if the frame rate is low or the original samples are not close in the time sequences. While these issues usually do not occur in supervised learning use cases, they could be pronounced during unsupervised domain adaptation. 
Because, in our experiments of UDA via self-training, mixing involves samples from distinct time sequences, such as two samples from distinct users or environments. This might produce infeasible joint positions, such as shortened or intersected arms. 
Further ablation studies might help understand this effect comprehensively.
Nevertheless, \oursys{} still achieved significant improvements in UDA. We think this could be attributed to the fact that the theoretical foundation of convex combination ~\cite{zhang2017mixup} can work well without requiring stringent preconditions.
Specifically, in convex combination, the mixing data augmentation encourages the model to behave linearly in between training examples. This linear behavior reduces the amount of undesirable oscillations in vicinal data distribution. Vicinal distribution can be noisy, for instance, traditional approaches even simply generate augmentation data from Gaussian noise.
For reference, ~\cite{zhang2017mixup} mixes images for classification, such as mixing images of dogs and cats pixel-wise, producing a new noisy image of an infeasible creature; but they also have shown the effectiveness of mixing-based data augmentation in improving data distribution and model performance. 
Furthermore, in future work, for both supervised learning and the UDA cases, we can adopt an add-on filter using forward kinematic constraints to alleviate this potential issue.}

\bibliographystyle{ACM-Reference-Format}
\bibliography{main}


\end{document}